\documentclass[12pt]{article}
\usepackage{a4wide}
\usepackage{graphicx}
\usepackage{amssymb}
\usepackage{amsmath}
\usepackage{slashed}
\usepackage{cite}
\usepackage{multicol}
\usepackage{braket}
\usepackage{hyperref}
\newcommand{\be}{\begin{equation}}
\newcommand{\ee}{\end{equation}}
\newcommand{\ba}{\begin{eqnarray}}
\newcommand{\ea}{\end{eqnarray}}

\begin{document}

%\twocolumn
\allowdisplaybreaks

\begin{titlepage}
\begin{flushright}
\end{flushright}
\vfill
\begin{center}
{\Large\bf Light vector dark matter with scalar mediator and muon g-2 anomaly}
\vfill
{\bf Karim Ghorbani}\\[1cm]
{Physics Department, Faculty of Sciences, Arak University, Arak 38156-8-8349, Iran}
%[.4cm]
%{$^b$Institute for Research in Fundamental Sciences (IPM)\\
% \it School of Particles and Accelerators,  \it P.O. Box 19395-5531, Tehran, Iran}

\end{center}
\vfill

\begin{abstract}
We study a model with a vector dark matter (DM) candidate interacting with the SM charged leptons 
through a scalar portal. The dark matter candidate acquires mass when the complex scalar 
breaks an abelian gauge symmetry spontaneously. The scalar interacts with the SM charged leptons 
through a dimension-6 operator. The scalar mediator induces elastic scattering of dark matter with electrons at tree level and also DM-nucleon interaction when 
the effects from scalar-Higgs mixing are also taken into account.
Given the recent results from Xenon1T upper bounds on DM-electron elastic scattering 
cross section where the strongest sensitivity lies in the range $\sim {\cal O}$(1) GeV, 
we find the viable space in the parameter space respecting constraints 
from the observed relic density, direct detection, muon $(g_\mu-2)$ anomaly, 
$e^+ e^-$ colliders, electron beam-dump experiments and astrophysical observables. 
It is shown that the current upper bounds of Xenon1T on DM-electron interaction is partially sensitive to the regions in the viable parameter space which is already excluded by the
electron beam-dump experiment, Orsay. 
We also find that there are viable DM particles with masses $\sim {\cal O}(1)$ GeV
evading the direct detection but stand well above the neutrino floor.
Almost the same viable regions are found when we apply the direct detection upper limits on 
the DM-proton spin-independent cross section.

\end{abstract}
\vfill
%keywords: Beyond the standard model, matter experiments, dark matter theory
%{\bf PACS:11.30.Rd, % Chiral symmetries
%          12.39.Fe, % Chiral Lagrangians
%          13.25.Jx, % Decays of other mesons
%          14.40.Aq. % pi, K, and eta mesons
% }
\vfill
%\footnoterule
{\footnotesize\noindent }

\end{titlepage}

%\begin{titlepage}
%\begin{flushright}
%\end{flushright}
%\vfill
%\begin{center}
%{\Large\bf Split fermionic WIMPs evade direct detection bounds}
%\vfill
%{\bf Karim Ghorbani and Parsa Hossein Ghorbani}\\[1cm]
%{Physics Department, Faculty of Sciences, Arak University, Arak 38156-8-8349, Iran}
%\end{center}
%\vfill

%\begin{abstract}
%\end{abstract}
%\vfill
%keywords: Beyond the standard model, dark matter experiments, dark matter theory
%{\bf PACS:11.30.Rd, % Chiral symmetries
%          12.39.Fe, % Chiral Lagrangians
%          13.25.Jx, % Decays of other mesons
%          14.40.Aq. % pi, K, and eta mesons
% }
%\vfill
%\footnoterule
%{\footnotesize\noindent }
%\end{titlepage}

\section{Introduction}
The nature of dark matter remains an unsolved problem and the solution
might reside at the intersection of cosmology and particle physics. 
The gravitational evidences for DM from cosmological observables are beyond doubt 
but its particle nature is still hypothetical \cite{Feng:2010gw,Bertone:2004pz}.
Weakly interacting massive particles (WIMPs) are vastly studied candidates for dark matter \cite{STEIGMAN1985375,Arcadi:2017kky,Bergstrom:2000pn}.
The mass of the wimpy dark matter can be very light as that for axions \cite{Kawasaki:2013ae}
or it may emerge at the TeV-scale \cite{PhysRevLett.64.615,Blum:2014dca}.  
The thermal production of DM in the early universe, known as the freeze-out process \cite{Steigman:2012nb}, is a natural paradigm
in cosmology resembling the same mechanism which makes very successful prediction
for light element abundance and cosmic microwave background.
The mass of the dark matter particle and its interaction type are key ingredients 
in searching for its direct detection (DD). Interactions with velocity suppressed or momentum suppressed
DM-nucleon scattering cross section are instances where DM may evade detection 
in direct and collider searches \cite{Nobile:2021qsn}.
Direct search for DM candidates with mass around 10 GeV up to hundred GeV 
having DM-nucleon interaction has been a dedicated strategy 
in underground DD experiments \cite{Akerib:2016vxi,Aprile:2017iyp,Agnese:2017njq}. 
In fact we do not know if WIMPs should necessarily interact with the atomic nuclei. 
At any rate, if DM interacts with nucleons it might be in the mass range that 
the current DD experiments cannot exclude it. This in turn advocates the absence of 
DM-nucleon elastic scattering in the current DD experiments to date.

One possible avenue in direct search for dark matter is that DM might 
interact exclusively with the SM leptons and possibly having suppressed interactions with nucleons. 
The focus here is on WIMP candidates with masses in the range $\lesssim 10$ GeV
communicating with the SM leptons by exchanging light scalar mediator. 
This type of interaction for DM receives stringent constraints 
from astrophysical and cosmological observations \cite{Cyburt:2015mya,Raffelt:1996wa,Ade:2015xua,Hinshaw:2012aka}.
Additionally, searches beyond the SM in rare kaon 
decays \cite{CortinaGil:2021gga,Krnjaic:2019rsv}, $e^+e^-$ colliders \cite{TheBABAR:2016rlg,Adachi:2019otg,BaBar:2020jma}, 
beam-dump experiments \cite{Davier:1989wz,Bjorken:1988as,Gninenko:2014pea,Chen:2017awl,Marsicano:2018vin} 
and muon anomalous magnetic moment (MAMM) \cite{Bennett:2006fi,Abi:2021gix} are highly 
motivated probes of light dark matter with leptophilic scalar mediator. 
Moreover, we apply the newest results from Xenon1T \cite{Aprile:2019xxb} which probe DM-electron scattering for DM masses in the range $(0.03-10)$ GeV.

The SM prediction for the muon magnetic moment reads 
$a_\mu^{\text{SM}} = (116591810 \pm 43) \times 10^{-11}$, 
where contributions from QED \cite{Aoyama:2012wk,Aoyama:2019ryr}, QCD or Lattice-QCD \cite{Davier:2017zfy,Keshavarzi:2018mgv,Colangelo:2018mtw,Hoferichter:2019mqg,Keshavarzi:2019abf,Kurz:2014wya,Davier:2019can,Borsanyi:2020mff,Melnikov:2003xd,Masjuan:2017tvw,Hoferichter:2018kwz,Gerardin:2019vio,Pauk:2014rta,Danilkin:2016hnh,Roig:2019reh,Blum:2019ugy} and electroweak interactions \cite{Czarnecki:2002nt,Gnendiger:2013pva} 
are taken into account with highest precision. 
The first measurement indicating a deviation form the SM prediction was found by Brookhaven
National Laboratory (BNL), $a_\mu^{\text{BNL}} = (116592089 \pm 63) \times 10^{-11}$ \cite{Bennett:2006fi}.
The newest measurement which confirms the deviation is announced by the Fermi National Laboratory (FNAL)
with improved statistics, $a_\mu^{\text{FNAL}} = (116592040 \pm 54) \times 10^{-11}$ \cite{Abi:2021gix}.
In order to explain the deviation a large number of investigations applying various models beyond the SM 
are performed. 
Among them there are models introducing DM candidates interacting with the SM leptons 
via leptophilic scalar \cite{Agrawal:2014ufa,Boehm:2020wbt,Garani:2019fpa,YaserAyazi:2019psw,Liu:2021mhn,Bai:2021bau,Ge:2021cjz,Horigome:2021qof,
Chun:2021dwx,Borah:2021jzu,Yin:2021mls}, via generic scalar mediator \cite{Athron:2021iuf,Arcadi:2021cwg,Zhu:2021vlz} 
and through vector mediator \cite{Bell:2014tta,Ghorbani:2017cey,Athron:2017drj,Arcadi:2021yyr}, 
emphasizing on the MAMM.

The present work examines a dark matter scenario in which the DM candidate is a vector gauge boson 
in an abelian scalar gauge theory. The gauge boson gets mass when the symmetry is broken spontaneously. 
Thus the mass of the gauge boson is confined by the gauge coupling and the vacuum 
expectation value (vev) of the new scalar. 
On the other side the scalar mediates the force between DM and the SM charged leptons. 
In this work the scalar interaction with the SM leptons is induced by dimension-6 operators.
The models with a scalar mediator motivated by an effective field theory with dimension-5 operators 
are studied in \cite{Batell:2017kty,Batell:2016ove,Chen:2015vqy}.
The main purpose of this work is two-fold. First, we would like to see if there can be found 
DM candidates and appropriate scalar mediators to explain the newest muon magnetic moment anomaly
and the same time satisfying other constraints from indirect searches. 
And finally we investigate to find out whether the most strongest upper limits 
on DM-electron and DM-proton scattering cross section from Xenon1T are sensitive to the remaining 
viable parameter space.

The structure of the paper is as follows. In section \ref{model} the DM model is presented and the effective
operators of dimension-6 are motivated by introducing a UV complete model.
Discussion on the evaluation of the DM abundance is given in section \ref{WIMP-Planck}.
A couple of different terrestrial and astrophysical constraints are introduced in section \ref{various}.
Our final results are shown in section \ref{final-results} after imposing the upper bounds from  
DD experiments. We will finish with a conclusion.

\section{Model}
\label{model}
The model we consider here contains a complex scalar field gauged under a $\text{U}^\prime$(1) symmetry with the Lagrangian
\begin{equation}
{\cal L}_{\text{DM}}   = (D_\mu \phi)(D^\mu \phi)^* - m^2 \phi \phi^* - \frac{1}{4} F'^{\mu\nu} F'_{\mu\nu}  \,,
\end{equation}
where $D_\mu = \partial_\mu - i g_\text{v} V_\mu$.  
The $\text{U}^\prime$(1) gauge symmetry is broken when the complex scalar field gets a non-zero vacuum expectation value, $v_s$. The scalar field can be parameterized as $\phi = \frac{1}{\sqrt{2}}(s+v_s) \exp(-i\pi/v_s)$. Here, $s$ and $\pi$ are real scalar fields. 
The Goldstone boson is "eaten" by the longitudinal component of the gauge field giving a mass to the 
gauge boson as $M_{V}= g_\text{v} v_s$.  

In addition, one may consider a type of low energy effective interaction for 
the complex scalar $\phi$ in the form of a dimension-6 operator as $\sim \frac{1}{\Lambda_l^2} |\phi|^2\bar{L} H l_R$. Here $H$ is the SM Higgs doublet, $L$ is the SM left-handed 
lepton doublet, $l_R$ is the right-handed SM lepton, and $\Lambda_l$ is an 
appropriate energy scale for lepton $l$. 
In principle the dimension-6 operators including the SM quarks like $\frac{1}{\Lambda_Q^2} |\phi|^2~\bar{Q} H^\dagger u_R$
and $\frac{1}{\Lambda_Q^2} |\phi|^2~\bar{Q} H d_R$ are allowed by the symmetry. 
These interactions induce a large contribution to the DM-nucleon elastic scattering 
leading to the exclusion of the entire parameter space by the current direct detection bounds. 
Through a UV complete model we will motivate a {\it lepton-specific} scenario in which only 
leptonic operator is important.

Here, we discuss a possible UV-completion of the above-mentioned effective interactions.
To this end, we introduce a heavy new Higgs doublet, $\Phi$, with appropriate quantum numbers.
The new doublet in general can have interactions with all the SM fermions. 
In this work we are interested in the so-called lepton-specific models in which the new doublet only
interacts with the SM leptons. This type of interaction for the new doublet is motivated in 
Two-Higgs doublet models \cite{Su:2009fz,Branco:2011iw,Marshall:2010qi}.
We consider the following UV model
\begin{equation}
 {\cal L}_{UV} = y_e \Phi \bar{L}_e  e_R + y_\mu \Phi \bar{L}_\mu  \mu_R + y_\tau \Phi \bar{L}_\tau \tau_R + \kappa \Phi^\dagger H |\phi|^2 + \text{h.c.}\,,
\end{equation}
where $H$ is the SM Higgs doublet. In the limit that the mass of the new doublet is heavy, integrating out the 
heavy doublet will lead us to the dimension-6 effective operator introduced earlier, i.e, $\sim \frac{1}{\Lambda_l^2} |\phi|^2~\bar{L} H l_R$.

Therefore, if we assume that the new scalar interacts with the SM particles only through the leptonic 
operator then it couples to the SM charged lepton currents effectively as, 
\begin{equation}
{\cal L}_{\text{eff}}   = \alpha_l s l^+ l^-   \,,
\end{equation}
where $l = e, \mu, \tau$, and $\alpha_l$ is the corresponding effective coupling constant. 
The effective couplings of the scalar to leptons are parameterized to be 
mass-hierarchical couplings, $\alpha_l = \frac{m_l}{v_s} c_l$, which is intriguing phenomenologically. 
The cutoff scale $\Lambda_l$ is obtained as $\Lambda_l^2\sim v_h v_s/\alpha_l$, where
$v_h$ is the vacuum expectation value of the SM Higgs.
Following the same line of reasoning in \cite{Batell:2017kty}, in the effective 
Lagrangian above we expect the two-loop contribution to the muon anomalous magnetic moment $(\Delta a_\mu)^{\text{2loop}}$ and the one-loop contribution to 
the muon anomalous magnetic moment $(\Delta a_\mu)^{\text{1loop}}$, 
satisfy the relation $(\Delta a_\mu)^{\text{2loop}}/(\Delta a_\mu)^{\text{1loop}} \sim \Lambda_\mu^2/(8\pi^2v_h^2)$. 
In order to have small two-loop contribution in comparison to the one-loop contribution we should
have $\Lambda_\mu < 2\sqrt2 \pi v_h \sim 2$ TeV.

The interaction Lagrangian which is relevant in this work includes these terms 
\begin{equation}
\label{int-Lag}
{\cal L}_{\text{int}}   = g_\text{v}^2 v_s  s V_\mu V^\mu + \frac{1}{2} g_\text{v}^2 s^2 V_\mu V^\mu 
 + \alpha_l s l^+ l^- + \frac{\alpha_l}{v_h} s h l^+ l^-  \,.
\end{equation}
The dark gauge boson $V$ is identified as our vector dark matter candidate.
In the rest of the paper we shall use $m_{V}$ and $m_{\text{DM}}$ exchangeably.
Moreover, we may consider another interaction term in the Lagrangian as 
$\sim \frac{1}{4} \lambda |\phi|^{2} H^\dagger H$, which can be arised from the potential part of the UV model. 
We will justify below that in order to respect bounds
from the invisible Higgs decay, the coupling $\lambda$ should be negligible.
This interaction is interesting here, because it cases mixing between the singlet scalar and the SM Higgs which can then lead to the invisible Higgs decay via the interaction $\sim (g_v^2 v_s \sin \theta) h V_\mu V^\mu$.
The mixing angle, $\theta$, which diagonalizes the scalar mass matrix  
satisfies the relation $\sin 2\theta = 2\lambda v_s v_h/(m_h^2 - m_s^2)$. 
The SM Higgs invisible decay width in the decay process $h \to VV$ is given by the formula
\begin{equation}
 \Gamma_{\text{inv}} = \frac{g_v^2v_s^2 m_h^3 \sin^2 \theta }{16\pi m_V^4} (1-4x^2+12x^4)(1-4x^2)^{1/2} \,,
\end{equation}
where $x = m_V/m_h$. 
The observed upper limit at 95\% confidence level on the branching ratio of the invisible Higgs decay is $\sim 0.19$ \cite{CMS:2018yfx}. 
Depending on the region of the parameter space we explore in this work, it is found that if the mixing angle $\theta$ lies 
in the range $\lesssim 8\times 10^{-4}$ then the respecting regions evade bounds from invisible Higgs decay. 

The last term in Eq.~\ref{int-Lag}, opens up the possibility of a new decay channel for the SM Higgs. The Higgs particle can then decay to a scalar $s$ in the process $h \to s \bar{f} f$, where $f$ stands for the SM leptons. In the following we present some results for the decay width of $h \to s \tau^+ \tau^-$, in terms of the scalar mass. We picked out this decay channel because it has the largest decay width among the others. 
The value $\alpha_\tau = 10^{-1}$ is chosen which is reasonably large enough to find 
the upmost contribution to the Higgs total decay width.
Since the decay width is proportional to $\alpha_\tau^2$, it is easy 
to estimate the decay width for other values of $\alpha_\tau$. 
To compute numerically the decay width $\Gamma(h \to s \tau^+ \tau^-)$ the 
code CalcHEP \cite{Belyaev:2012qa} is employed. 
\begin{table}[h]
 \caption{The decay width $\Gamma(h \to s \tau^+ \tau^-)$. The relevant effective coupling $\alpha_\tau = 10^{-1}$.}
 \begin{center}
 \begin{tabular}{c rrrrrrr}
 \hline\hline
  $m_s$ [GeV]    & $10^{-3}$    & 0.1        & 1          & 5         & 10         & 50         & 100 \\
 \hline
 $\Gamma$~[$10^{-5}$GeV]&$6.68$&$5.76$&$5.16$&$4.5$&$3.90$&$0.675$&$0.002.49$   \\
 \hline
 \end{tabular}
 \end{center}
 \label{decay-width}
 \end{table}
Our results for the decay width is presented in Table~\ref{decay-width}. 
We can estimate that the total decay width, $\Gamma(h \to s \bar{f} f)$, for the scalar 
mass of interest in this work is of order $\sim 10^{-5}$ GeV. 
The total decay width of the SM Higgs is $3.2^{+2.8}_{-2.2}$ MeV \cite{ParticleDataGroup:2020ssz}.  
In conclusion, the total decay width $\Gamma(h \to s \bar{f} f)$ is about two orders of magnitude smaller than 
the Higgs total decay width and therefore the measured Higgs decay width will not put any constraint 
on the relevant parameters.

\section{Constraints from WMAP/Planck observation}
\label{WIMP-Planck}
In light of lacking any evidence of 
hundred GeV DM in direct detection searches so far, the interest has pushed towards 
low mass DM or light DM with $\sim$ GeV DM particles.
In this work we adopt thermal production of light DM particles through 
the so-called freeze-out mechanism which sounds natural and regarded as an standard mechanism for thermal relic.
During this thermal process DM annihilation to the SM particles (visibles) or 
other particles (secluded sector), and the reverse processes take place. 
The annihilation rate is in competition with the expansion rate of the Universe in the early universe. 
There is a special temperature called freeze-out temperature, $T_{f}$,
or decoupling temperature around which the DM 
particles get out of equilibrium and its density remains constant thereafter. 
The stronger the DM interaction with the SM particles, the longer it takes for DM particles to freeze out.
Dark matter relic density is a function of the thermally averaged annihilation cross section, $\langle \sigma v \rangle$, as $\Omega h^2 \propto \langle \sigma v \rangle^{-1}$.
The observed value of the dark matter density is $\Omega h^2 \approx 0.12$ \cite{Hinshaw:2012aka,Ade:2015xua}.
The theoretical value for the DM relic density in the model parameter space is obtained 
by solving the relevant Boltzmann equation numerically applying 
micrOMEGAs \cite{Belanger:2013oya}.

Initially, we would like to find viable regions in the parameter space with DM masses 
in the range $10^{-3}~\text{GeV} < m_{\text{DM}} < 10~\text{GeV}$ which give 
rise to a relic abundance consistent with the observed value provided by 
WMAP \cite{Hinshaw:2012aka} and Planck \cite{Ade:2015xua}. 
When thermal WIMPs have s-wave $2\to2$ annihilation to visible final states, observed DM density
puts a lower limit on the WIMP mass, $m \gtrsim$ 20 GeV \cite{Leane:2018kjk}. 
However, this is not the case when WIMPs also annihilate to secluded dark sector.
There are two possible DM annihilation channels for the DM particles in our model, namely, annihilation to a pair of dark scalars and annihilation to a pair of the SM charged leptons. In fact, we stay in a region of parameter space that $2 \to 2$ annihilation processes are dominant. 
DM particles can annihilate as $VV \to ss$ via $t$- and $u$-channels with exchanging a 
vector boson and annihilation to a pair of dark scalar through a contact 
interaction, if $m_{\text{DM}} > m_s$. 
In addition, the s-channel DM annihilation, $VV \to s \to e^+ e^-, \mu^+ \mu^-, \tau^+ \tau^-$ 
will be accessible when kinematically allowed. When we consider the scalar-Higgs mixing processes
another annihilation channel, $V V \to h \to f^+ f^-$, becomes possible. However, since its 
annihilation cross section is proportional to $\sin^2 \theta$ and $\theta$ is restricted to quite small values, this effect has a very small contribution to the DM relic density.  
The relevant Feynman diagrams for the DM annihilation processes with dominant contributions
are depicted in Fig.~\ref{feynman-anni}. 
The analytical formulas for the DM annihilation cross sections are given in the Appendix.  
The analytical results are confirmed after implementing our model in the code CalcHEP \cite{Belyaev:2012qa}. 

\begin{figure}
\begin{center}
\includegraphics[width=0.65\textwidth,angle =0]{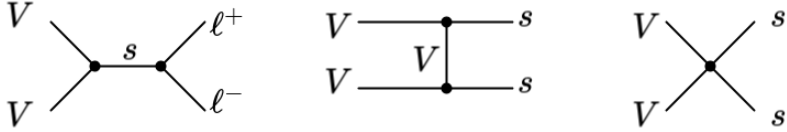}
\end{center}
\caption{Feynman diagrams for DM annihilation with dominant contributions are shown.}
\label{feynman-anni}
\end{figure}

\begin{figure}
\hspace{-.8cm}
\begin{minipage}{0.37\textwidth}
\includegraphics[width=\textwidth,angle =-90]{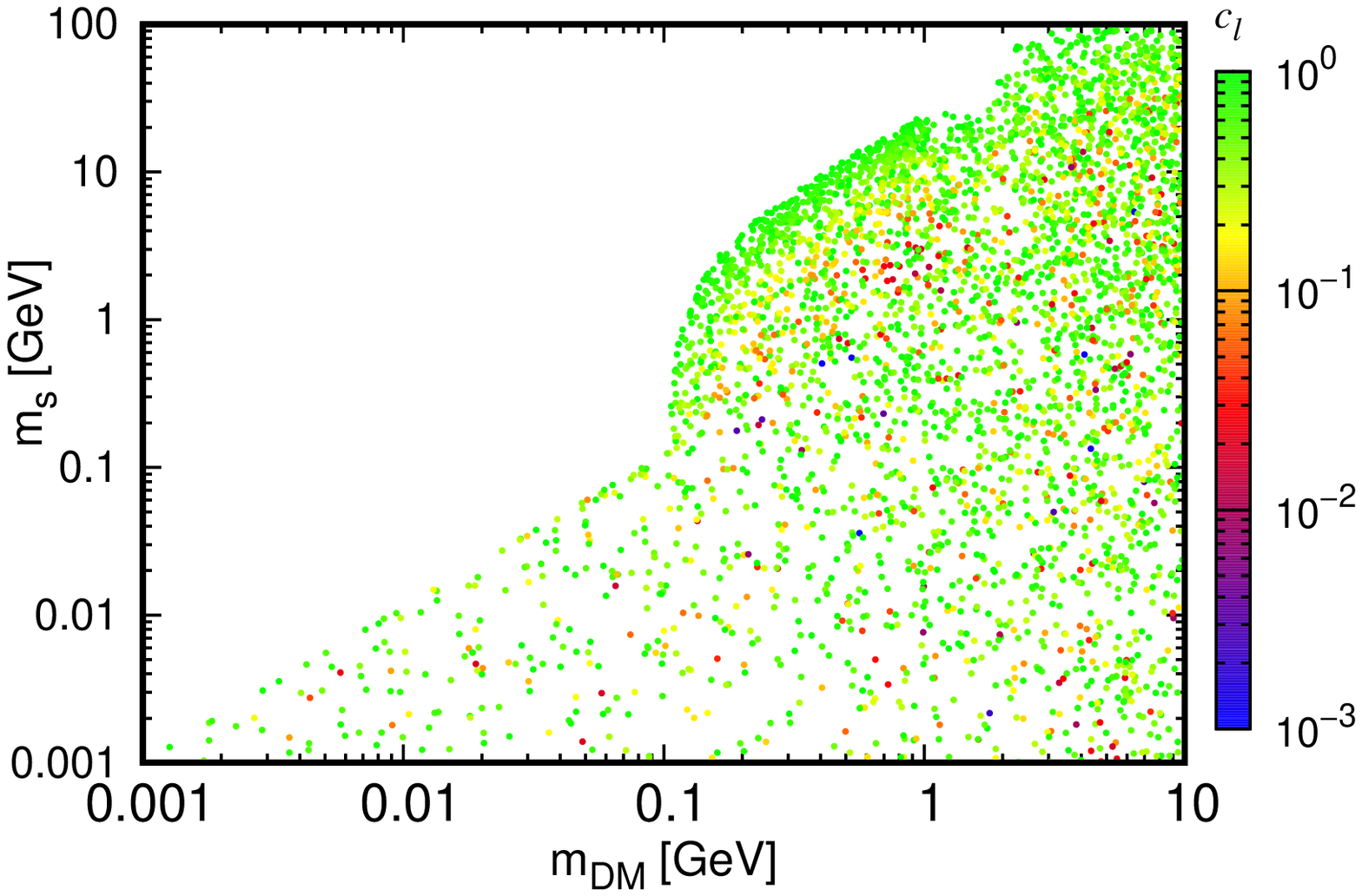}
\end{minipage}
\hspace{3cm}
\begin{minipage}{0.37\textwidth}
\includegraphics[width=\textwidth,angle =-90]{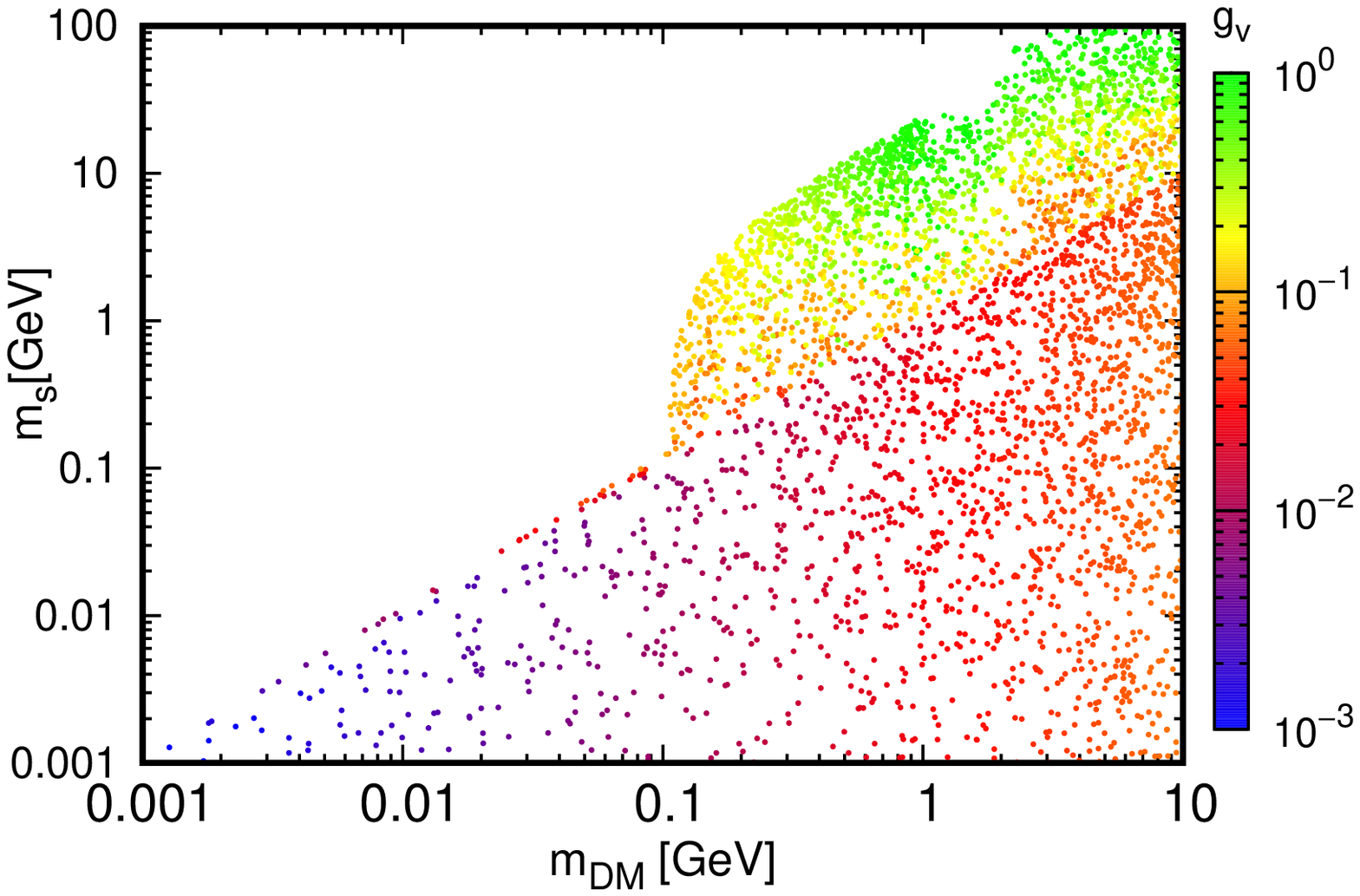}
\end{minipage}
\begin{center}
\includegraphics[width=0.37\textwidth,angle =-90]{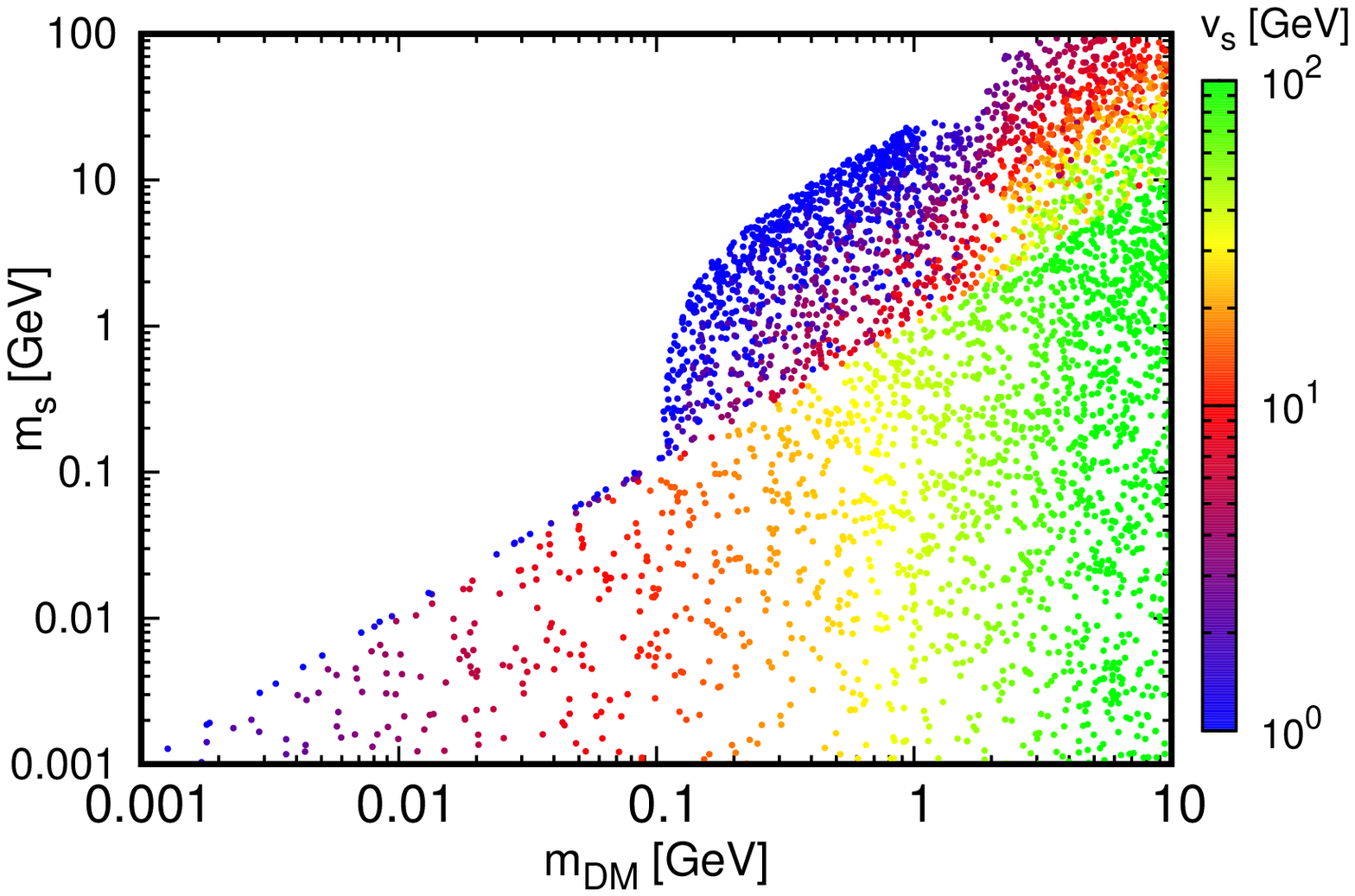}
\end{center}
\caption{In these plots we only applied the constraints on the relic density from WMAP/Planck.
In the plane $m_\text{DM}-m_s$, viable ranges for the parameters 
$m_\text{DM}$, $m_s$, $c_l$, $g_\text{v}$ and $v_s$ are shown in three plots.}
\label{relic}
\end{figure}

The parameter space we scan over, lies in the following intervals: $10^{-3}~\text{GeV}<m_s<100~\text{GeV}$, $0<g_\text{v}<1$, $0< c_e = c_\mu = c_\tau <1$ and 
$1~\text{GeV}<v_s<300~\text{GeV}$.
Lets remind that $m_{V} = m_{\text{DM}} = g_\text{v} v_s$ and $\alpha_l = (\frac{m_l}{v_s}) c_l$.
In our scan the number of sampling is $10^7$.
In each sampling only when the computed relic density is consistent with the observed DM relic density
we keep the sampled free parameters. After finding the viable values for the 
parameters, $c_l$, $g_\text{v}$, $v_s$, $m_s$ and $m_\text{DM}$, we present in 
the plane $m_{\text{DM}}-m_s$ the resulting 
values for $c_l$, $g_\text{v}$ and $v_s$ in three plots, respectively in Fig.~\ref{relic}.
It is evident from the results shown in Fig.~\ref{relic} that larger DM mass towards 10 GeV 
requires larger mass for the scalar up to about 100 GeV.

\section{Various constraints on scalar-muon coupling}
\label{various}
In this section we discuss several types of constraints that might
affect the viable parameter space.

 I)  {\it Muon anomalous magnetic moment} 

The precise measurement of the muon anomalous magnetic moment, $a_{\mu}$, has been under intense 
scrutiny since long time, for a review on this regards one may consult \cite{Aoyama:2020ynm}.
This quantity is defined as $a_{\mu} = \frac{g_{\mu}-2}{2}$, where $g_{\mu}$ is the well-known 
gyromagnetic ratio for muon. At tree level in perturbation theory the quantity $g_{\mu}$ 
reads, $g_{\mu} =2$. 
The SM radiative corrections include loop contributions from QED, QCD and weak interactions.       
The theoretical prediction of muon magnetic moment in the SM is suffered mainly 
from the uncertainties in the hadronic vacuum polarization and the hadronic light-by-light scattering.
A sizable deviation, $\Delta a_{\mu}$, observed in the past experiments 
at the Brookhaven National Laboratory (BNL) experiments \cite{Bennett:2006fi} considered as 
a footprint of a probable new physics, taking into account the controllable uncertainties 
from the theoretical side.
The new updated data from the muon g-2 experiment at Fermi National Laboratory (FNAL) not 
only supports the long-standing discrepancy but also provides results with improved 
statistics \cite{Abi:2021gix}.
The new result comes along with a significance of about 4.2 sigma and indicates a positive 
excess over the SM prediction. An updated experimental world average gives   
$\Delta a_{\mu} = a_{\mu}^{exp}-a_{\mu}^{\text{SM}} = (2.51 \pm 0.59) \times 10^{-9}$.

As a new physics effect, the scalar mediator in the present model will contribute  
to the muon anomalous magnetic moment at loop level and leads to the correction, 
\begin{equation}
\begin{aligned}
 \Delta a_{\mu}^{\text{NP}} & = \frac{\alpha_\mu^2}{8\pi^2} \int_{0}^{1} \frac{(1-z)^2 (1+z)}{(1-z)^2+b^2 z} dz
                \\ &
                =
 \frac{\alpha_\mu^2}{8\pi^2} \Big[ \frac{1}{2} \left(-2 b^2+\left(b^2-3\right) b^2 \log \left(b^2\right)-2 \sqrt{b^2-4} \left(b^2-1\right) b \tanh ^{-1}\left(\frac{b^2-2}{b \sqrt{b^2-4}}\right)+3\right)
  \\ &
  + \frac{b \left(b^4-5 b^2+4\right) \tanh ^{-1}\left(\frac{b}{\sqrt{b^2-4}}\right)}{\sqrt{b^2-4}} 
  \Big] \,,
\end{aligned}
\end{equation}
where $b = \frac{m_s}{m_\mu}$. 
The new data on the muon magnetic moment deviation puts stringent constraints on
the scalar-muon coupling and the scalar mass. \\

II) {\it $e^+ e^-$ annihilation in colliders}

In $e^+ e^-$ colliders the production of the new scalar is possible 
through the process $e^+ e^- \to \mu^+ \mu^- s$. The scalar will subsequently
decay to $\mu^+ \mu^-$ and therefore there are 4 muons in the final state. 
The BaBar experiment has done search in this channel and
found the strongest upper limits on the effective coupling, $\alpha_\mu$, for 
$m_s > 2m_\mu$ \cite{TheBABAR:2016rlg}.
For scalar masses with $m_s < 2 m_\mu$, the Belle II experiment found constraints
in search for scalar production in the same channel but with the subsequent
decay $s\to \text{Invisible}$ \cite{Adachi:2019otg}.
Moreover, the BaBar experiment has found constraints on a leptophilic scalar ($\Phi_L$) decaying predominantly to leptons \cite{BaBar:2020jma}. The limits constrain the scalar coupling for scalar masses 
up to $\sim 7$ GeV.
\\

III) {\it Beam-dump experiments}
   
   Proton and electron beam-dump experiments are suitable probes in search for 
     new physics at low energy. In particular, a secondary muon beam originated from 
     the original beam may radiate a dark scalar and the scalar can decay subsequently into the SM      leptons. Therefore, it is possible to search for the scalar-lepton coupling 
     in these experiments. We apply exclusion limits on the scalar-muon coupling 
     from two electron beam-dump experiments, Orsay \cite{Davier:1989wz} and E137 \cite{Bjorken:1988as}.

IV) {\it Meson decays}

Since the scalar mediator, $s$, in our model is leptophilic 
the meson decays as $B\to K s$ and $K\to \pi s$ are not possible and there exist
no constraints in this regard.\\

V) {\it Supernova cooling}

The stellar cooling processes such as supernova cooling
are type of probes which are sensitive to scalar-muon 
coupling for the scalar masses below $\sim 1$ MeV \cite{Bollig:2020xdr}.
In this work the interest is mainly in the scalar mass $\gtrsim 1$ MeV.\\

VI) {\it BBN }

BBN put constraints on the effective number of relativistic degrees of freedom
beyond the SM particles with $\Delta N_{\text{eff}} \lesssim 0.2-0.6$ \cite{Mangano:2011ar}.
In case our new particles have mass $\gtrsim {\cal O}$(1) MeV then  
the parameter space is not sensitive to the BBN bounds \cite{Pospelov:2010hj}. 

\begin{figure}
\begin{center}
\includegraphics[width=0.6\textwidth,angle =0]{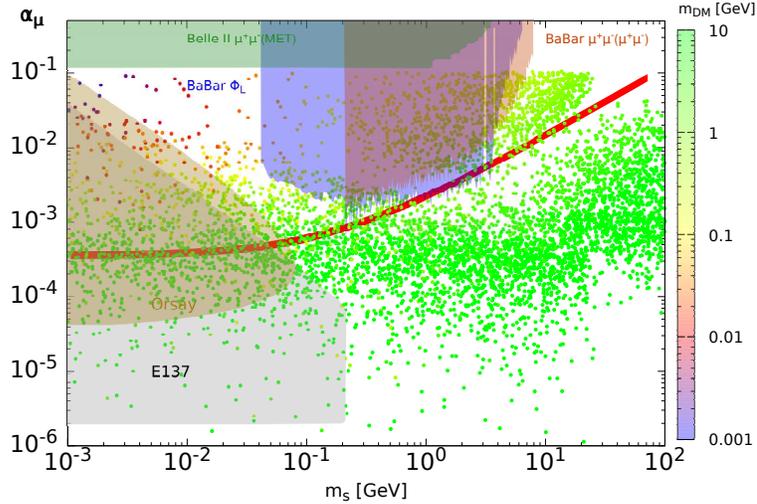}
\end{center}
\caption{The points in color show the viable space respecting 
the observed DM abundance in the place $\alpha_\mu$-$m_s$. 
Constraints from $e^+e^-$ colliders and
electron beam-dump experiments are imposed. 
The allowed region from muon anomalous magnetic moment is also shown as a red band.}
\label{exclusion}
\end{figure}

Taking into account all the constraints mentioned in this section and that from observed relic density 
we show the viable region of the parameter space in Fig.~\ref{exclusion}. 
The supernova cooling constraints exclude the allowed region by ($g_\mu-2$) anomaly 
for scalar masses smaller than $\sim$ 1 MeV, while the BaBar (in the 
process $e^+ e^- \to \mu^+ \mu^- (\mu^+ \mu^-)$) 
and Belle II upper limits do not overlap with the allowed region. 
However, the limits from BaBar (in the process $e^+ e^- \to \tau^+ \tau^- \Phi_L$)
partially exclude the allowed region in the scalar mass range $\sim 1~\text{GeV}- 4~\text{GeV}$.
It is also seen that in the remaining parameter space 
respecting the $g_\mu-2$ allowed region, the observed 
relic density and Beam-dump experiments, DM mass varies in the 
range $\sim 0.1~\text{GeV}-10~\text{GeV}$ and the scalar mass in the 
range $\sim 0.07~\text{GeV}-20$ GeV. 
The strongest lower limits on the scalar mass belongs to two electron beam-dump experiments where scalar 
masses smaller than $\sim 0.07$ GeV are excluded by the Orsay experiment.

\section{Direct detection bounds}
\label{final-results}
\begin{figure}
\begin{center}
\includegraphics[width=0.35\textwidth,angle =0]{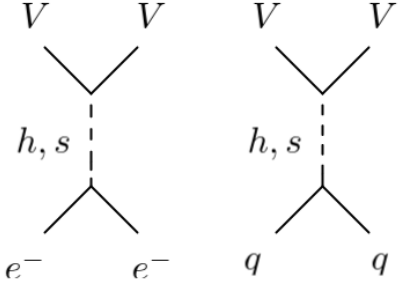}
\caption{Feynman diagrams for DM-electron and DM-quark elastic scattering at 
tree level.}
\end{center}
\label{diagramsDD}
\end{figure}

In our model spin-independent (SI) DM-nucleon interaction is present at tree level, due to the scalar-Higgs mixing effects. In addition, DM-electron elastic scattering 
of spin-independent type is feasible at tree level. 
In the following we ignore the loop suppressed DM-matter interactions. 
In Fig.~\ref{diagramsDD} Feynman diagrams for the DM-electron scattering at tree level
and also DM-quark scattering at tree level are depicted.
We obtain a reference DM-electron direct detection cross section, 
\begin{equation}
\sigma^{e}  \sim  \frac{4}{3\pi} \alpha_e^2 g_{\text{v}}^2 \frac{\mu_{\text{ve}}^2}{(m_s^2+\alpha^2m_e^2)^2} \,,
\end{equation}
where the reduced mass of the DM-electron is $\mu_{\text{ve}}$ and the electron momentum transfer is typically set by $q\sim \alpha m_e$. The contribution of the diagram with the Higgs propagator
to the DM-electron cross section is numerically negligible since the mixing angle, $\theta$, is very small.
In the limit that $m_s \gg \alpha m_e$, the DM form-factor $F_{\text{DM}}\sim 1$  \cite{Essig:2011nj}. 
So far, in direct detection experiments there is found no evidence of DM-electron 
elastic scattering. However, recently the experimental results from 
Xenon10 \cite{Essig:2011nj}, DarkSide-50 \cite{Agnes:2018oej} 
set upper bounds on DM-electron for masses below $\sim$ 1 GeV 
and Xenon1T \cite{Aprile:2019xxb} provides stringent bounds on the DM-electron cross 
section for DM masses in the range $\sim 0.03- 10$ GeV \cite{Aprile:2019xxb}.  
On the other hand, the neutrino floor sets the lowest limits for the scattering cross section of dark matter with visible matter. We apply the latest result for the neutrino floor given in \cite{Billard:2021uyg}.
\begin{figure}
\hspace{-1.cm}
\begin{minipage}{0.37\textwidth}
\includegraphics[width=\textwidth,angle =-90]{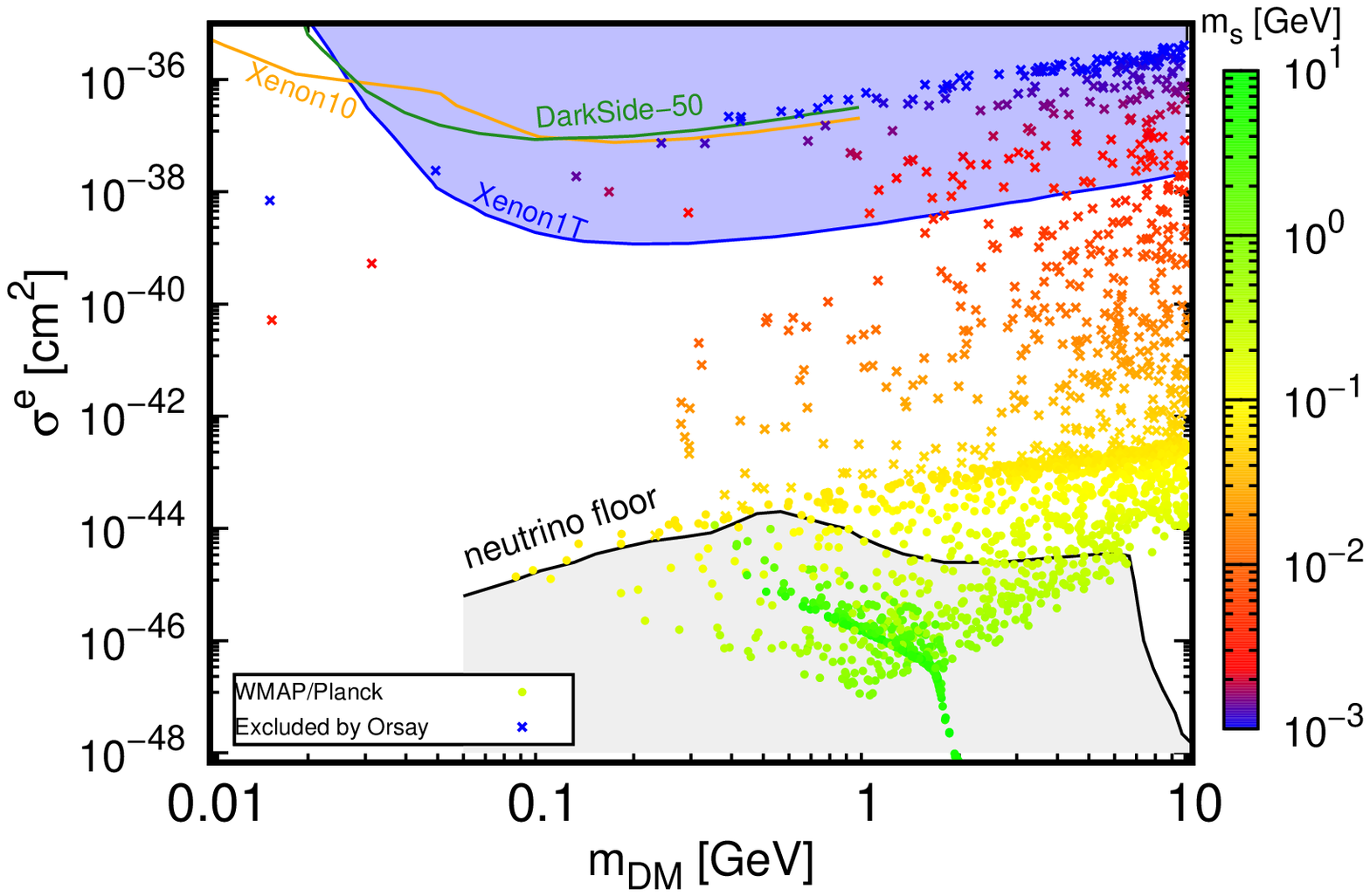}
\end{minipage}
\hspace{2.7cm}
\begin{minipage}{0.37\textwidth}
\includegraphics[width=\textwidth,angle =-90]{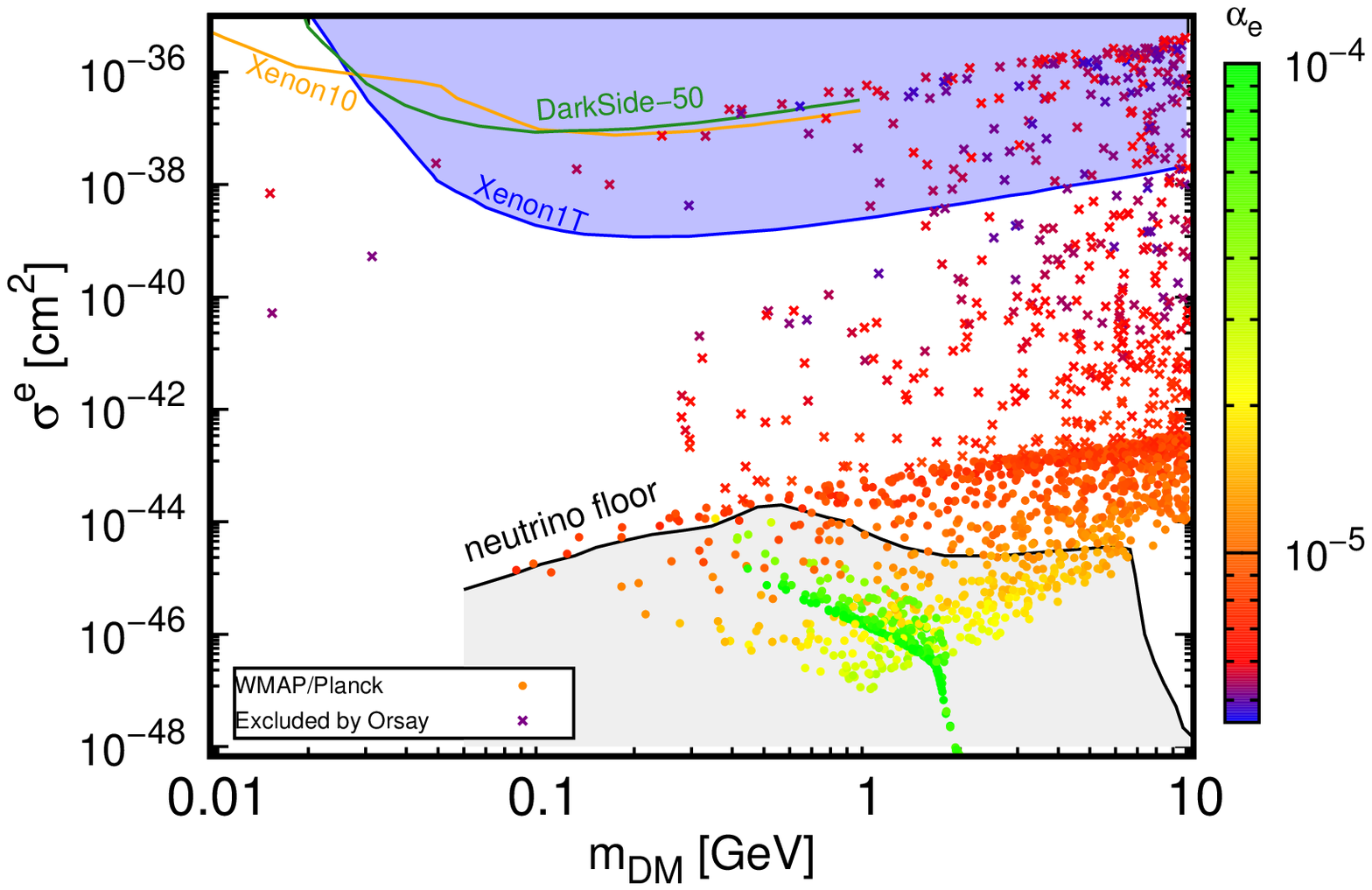}
\end{minipage}
\caption{We show regions in the parameter space which respect all the constraints discussed in the text and also points which are excluded by the electron beam-dump experiment, Orsay. All the points respect the allowed region by the muon $(g_\mu-2)$ anomaly. 
The upper bounds from direct detection experiments on the DM-electron elastic cross section are shown. In the left panel the scalar mass, $m_s$, and in the right panel the coupling, $\alpha_e$,
are shown in the vertical color spectrum. The neutrino floor is shaded in gray.}
\label{direct-electron}
\end{figure}
\begin{figure}
\begin{center}
\includegraphics[width=0.45\textwidth,angle =-90]{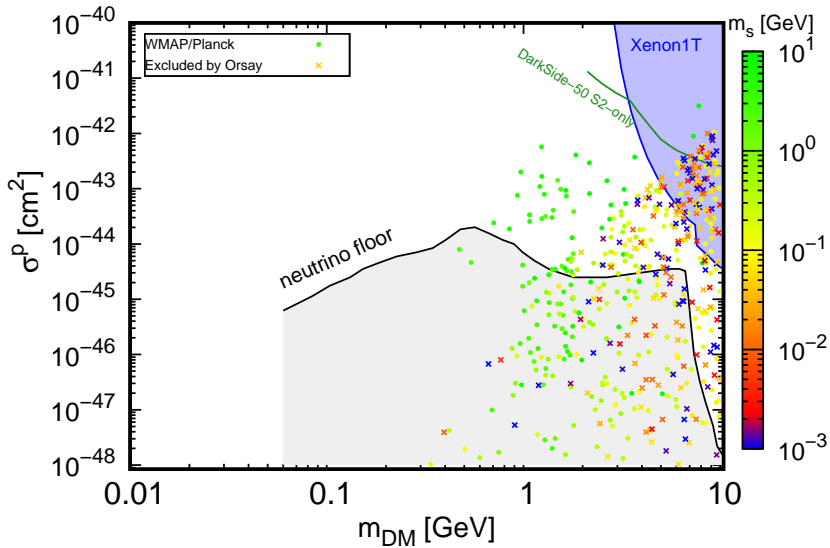}
\end{center}
\caption{In this plot we show regions in the parameter space which respect all the constraints discussed in the text and also points which are excluded by the electron beam-dump experiment, Orsay. All the points respect the allowed region by the muon $(g_\mu-2)$ anomaly. 
The upper bounds from direct detection experiments on DM-proton elastic cross section 
are shown. The viable values for the scalar mass, $m_s$, is also shown. The neutrino floor is shaded in gray.}
\label{direct-nucleon}
\end{figure}
In this section we pick out points in the parameter space which respect all the relevant 
constraints discussed previously, including those from beam-dump experiments, observed relic density.
We also confine the parameter space to the regions allowed by the muon $(g_\mu-2)$ anomaly.  
The regions in the parameter space that we scan over are: $0 < g_\text{v} < 1$, $1~\text{GeV} < v_s < 300~\text{GeV}$, $0 < c_l < 1$ 
and $10^{-3}~\text{GeV} < m_s < 10~\text{GeV}$. For the DM mass we have $m_V = g_\text{v} \times v_s$, 
and the relevant effective coupling here is $\alpha_e = \frac{m_e}{v_s} c_e$. 
The final result for DM-electron elastic scattering cross section in terms of 
the DM mass and the scalar mass, $m_s$, is shown in left panel of Fig.~\ref{direct-electron}
and in terms of DM mass and the coupling, $\alpha_e$, is presented in the right panel of
Fig.~\ref{direct-electron}. 
The results indicate that Xenon1T having the strongest limits among other DD experiments 
is only sensitive to the region with the scalar mass which is already 
excluded by the electron beam-dump experiments, Orsay.
However, there are regions with $m_s \gtrsim 0.07$ GeV and with dark matter mass in the 
range $1~\text{GeV} \lesssim m_{\text{DM}} \lesssim 10$ GeV which evade 
the current Xenon1T bounds and stand well above the neutrino floor.

Xenon1T \cite{Aprile:2019xxb} and DarkSide-50 \cite{Agnes:2018oej} collaborations provide bounds on the DM-nucleon cross section for DM masses below 10 GeV, as shown in Fig.~\ref{direct-nucleon}. We apply the package 
micrOMEGAs to compute the DM-proton SI cross section in the parameter space in 
the same ranges we discussed on the DM-electron case. 
Concerning the mixing angle, $\theta$, we always pick values to respect 
the invisible Higgs decay bounds. We show our results in Fig.~\ref{direct-nucleon} for points 
which respect all the restrictions and also points which are excluded by the 
Orsay beam-dump experiment. We find that there are viable DM candidates with masses $\sim 0.7- 10$ GeV and scalar mass $m_s \sim 0.1-10$ GeV with SI cross section well above the neutrino floor and respecting the available DD bounds.

\section{Conclusion}
In light of the newest results from the muon magnetic moment anomaly, ($g_\mu-2$), and 
DM-matter elastic scattering upper bounds from Xenon-1T, we exemplified a vector 
DM model with a scalar mediator which is coupled to the SM charged leptons via dimension-6 operators. 
We introduced a UV complete model to motivate the types of dimension-6 operators used in our study.
From phenomenological point of view, we confined the dark matter mass to 
the range $ 10^{-3}~\text{GeV} < m_{\text{DM}} < 10$ GeV and the scalar masses in the range 
 $ 10^{-3}~\text{GeV} < m_s < 100$ GeV.

In the first part of the analysis we imposed constraints from the observed DM density, muon anomalous magnetic moment, supernova cooling, $e^+e^-$ colliders and electron beam-dump experiments. 
The viable range for the scalar mass is then obtained as 
$ 0.07~\text{GeV} \lesssim m_{s} \lesssim 20$ GeV 
and for the DM mass as $0.1~\text{GeV} \lesssim m_{\text{DM}} \lesssim 10$ GeV.
Next we computed the DM-electron elastic scattering cross section. 
We then apply the upper limits from the DD experiments, Xenon100, DarkSide and Xenon1T
and find that the strongest bound from Xenon1T excludes scalar masses with $m_{s} \lesssim 3$ MeV
for DM masses $0.1~\text{GeV} \lesssim m_{\text{DM}} \lesssim 10$ GeV.
Since we already had found that electron beam-dump experiment, Orsay, excludes scalar masses with 
$m_s < 0.07$ GeV, we can conclude that the current DD experiments via DM-electron interaction have almost two orders of magnitude weaker sensitivity reach on the scalar mass than the electron beam-dump experiments. Given that the neutrino floor is increasing in the region with DM mass smaller than 10 GeV, we are still able to find DM candidates of ${\cal O}(1)$ GeV with direct detection cross section about two orders of magnitude above the neutrino floor.

Moreover, considering the DM-nucleon interaction for DM mass below 10 GeV, viable regions are 
found that are not yet explored by the DD experiments and further improvements on the experimental 
bounds in this mass range would be essential in order to further constrain or exclude the dark matter models.     

\section{Acknowledgment}
The author would like to thank Dr. Parsa Ghorbani for useful discussions.

\section{Appendix: Annihilation cross sections}
\label{Apen}
Here we present the formulas for the DM annihilation cross section times the relative velocity.
First, the annihilation cross section for the $s$-channel
annihilation process $V V \to l^+ l^-$ with $l = e, \mu, \tau$, is obtained  
\begin{equation}
 \sigma_{\text{anni}} v_{rel} (V V \to l^+ l^-) = \frac{2 \alpha_l^2 v_s^2 g_{\text{v}}^4}{9\pi^2}  \frac{(1-4m_l^2/s)^{3/2}}{(s-m^2_s)^2} \,.
  \end{equation}
And then we find the DM annihilation cross section with a pair of singlet scalars in the final state
\begin{equation}
\begin{aligned}
 \sigma_{\text{anni}} v_{rel} (V V \to s s) & = \frac{\sqrt{1-4m^2_s/s}}{16\pi^2s}
\int d\Omega \Big[ \frac{64}{9} v_s^4 g_{\text{v}}^8 \Big(\frac{1}{t-m_V^2} + \frac{1}{u-m_V^2}\Big)^2
  \\ &
  -\frac{64}{9} v_s^2 g_{\text{v}}^6 \Big(\frac{1}{t-m_V^2}+\frac{1}{u-m_V^2}\Big) + \frac{8}{9} g_{\text{v}}^4 \Big] \,,
\end{aligned}
\end{equation}
where in the formulas above, $s$, $t$ and $u$ are the relevant mandelstam variables. The relative velocity of the incoming DM particles is denoted by $v_{rel}$.

\bibliography{ref}

\bibliographystyle{utphys}

\end{document}